\documentclass[twocolumn,showpacs,superscriptaddress,preprintnumbers,amsmath,amssymb,pre]{revtex4}

\usepackage{graphicx}
\usepackage{bm}
\usepackage{times}

\begin{document}



\newcommand{\dd}{\mathrm{d}}
\newcommand{\gcc}{\mbox{g~cm$^{-3}$}}
\newcommand{\kB}{k_\mathrm{B}}
\newcommand{\etal}{{et al.}}
\newcommand{\beq}{\begin{equation}}
\newcommand{\eeq}{\end{equation}}
\newcommand{\bea}{\begin{eqnarray}}
\newcommand{\eea}{\end{eqnarray}}
\newcommand{\req}[1]{Eq.\ (\ref{#1})}
\newcommand{\Ftot}{F_\mathrm{tot}}
\newcommand{\FC}{F_{\mathrm{Coul}}}
\newcommand{\Nion}{N_i}


\newcommand{\aap}[1]{{Astron.\ Astrophys.} \textbf{#1}}
\newcommand{\ApJ}[1]{{Astrophys.\ J.} \textbf{#1}}
\newcommand{\ApJS}[1]{{Astrophys.\ J. Suppl.\ Ser.} \textbf{#1}}
\newcommand{\CPP}[1]{{Contrib.\ Plasma\ Phys.} \textbf{#1}}
\newcommand{\HEDP}[1]{{High Energy Density Phys.} \textbf{#1}}
\newcommand{\JPA}[1]{{J.\ Phys. A } \textbf{#1}}
\newcommand{\JPB}[1]{{J.\ Phys. B } \textbf{#1}}
\newcommand{\JETP}[1]{{JETP} \textbf{#1}}
\newcommand{\JQSRT}[1]{{J.~Quant.\ Spectrosc.\ Radiat.\ Transfer} \textbf{#1}}
\newcommand{\MNRAS}[1]{{Mon.\ Not.\ R.\ Astron.\ Soc.} \textbf{#1}}
\newcommand{\PoP}[1]{{Phys.\ Plasmas} \textbf{#1}}
\newcommand{\PR}[1]{{Phys.\ Rev.} \textbf{#1}}
\newcommand{\PRA}[1]{{Phys.\ Rev. A} \textbf{#1}}
\newcommand{\PRB}[1]{{Phys.\ Rev. B} \textbf{#1}}
\newcommand{\PRE}[1]{{Phys.\ Rev. E} \textbf{#1}}
\newcommand{\PRL}[1]{{Phys.\ Rev.\ Lett.} \textbf{#1}}


\title{Equation of state for partially ionized carbon 
and oxygen mixtures at high temperatures}

\author{G\'erard Massacrier}\email{Gerard.Massacrier@ens-lyon.fr}
    \affiliation{Universit\'e de Lyon, Lyon F-69003; 
    Universit\'e Lyon~I, 
    Observatoire de Lyon,  Saint-Genis-Laval F-69230}
    \affiliation{Ecole Normale Sup\'erieure de Lyon, CNRS, UMR~5574, 
    Centre de Recherche Astrophysique de Lyon,
    F-69007, France}
\author{Alexander Y. Potekhin}\email{palex@astro.ioffe.ru}%
    \affiliation{Universit\'e de Lyon, Lyon F-69003; 
    Universit\'e Lyon~I, 
    Observatoire de Lyon,  Saint-Genis-Laval F-69230}
    \affiliation{Ecole Normale Sup\'erieure de Lyon, CNRS, UMR~5574, 
    Centre de Recherche Astrophysique de Lyon,
    F-69007, France}
    \affiliation{Ioffe Physical-Technical Institute
    of the Russian Academy of Sciences,
     194021 St.\ Petersburg, Russia}
\author{Gilles Chabrier}\email{chabrier@ens-lyon.fr}
    \affiliation{Ecole Normale Sup\'erieure de Lyon, CNRS, UMR~5574, 
    Centre de Recherche Astrophysique de Lyon,
    F-69007, France}


\date{\today}

\begin{abstract}

The equation of state (EOS) for partially ionized carbon, oxygen, and
carbon-oxygen mixtures 
at temperatures  
$3\times10^5\mbox{~K}\lesssim T\lesssim 3\times10^6 $~K 
is calculated  over a wide range of densities, using the method of
free energy minimization in the framework of the chemical picture
of plasmas.  The free energy model is an improved extension
of our model previously developed for pure carbon
(\PRE{72}, 046402). The internal partition functions of bound species
are calculated by a self-consistent  treatment of each ionization
stage in the plasma environment taking into account pressure
ionization. The long-range Coulomb interactions between ions and
screening of the ions by free electrons are included using our
previously published analytical model, recently
improved, in particular for the case of mixtures. We also propose
a simple but accurate method of calculation of the EOS of
partially ionized binary mixtures based on detailed ionization
balance calculations for pure substances.

\end{abstract}

\pacs{52.25.Kn, 05.70.Ce, 52.27.Gr}


\maketitle

\section{Introduction}

An understanding of the physical properties of matter at high
densities and temperatures is important as a problem of fundamental
physics as well as  for various applications. A particularly
challenging problem is the calculation of the equation of state
(EOS) for stellar partial  ionization zones, where the electrons
and the different ionic species cannot be regarded as simple mixtures
of ideal gases:  Coulomb interactions, bound-state level shifts,
pressure ionization, and electron degeneracy must be taken into
account.  In a previous publication \cite{PMC05}, we calculated
the EOS for carbon at temperatures $10^5\mbox{~K}\lesssim T
\lesssim 10^7 $~K  over a wide range of densities $\rho$, based on
the free energy minimization method \citep{Harris-ea60,GHR69}
which enables us to include the complex physics in the model and
ensures thermodynamic consistency. For the case of carbon, in particular, the EOS
developed by Fontaine et al.~\cite{FGV77} (FGV) more than three
decades ago and widely used in astrophysics up to now is based on
the free energy minimization method at relatively low densities
$\rho\lesssim0.1$ \gcc.

The free energy model inevitably
becomes complicated when density increases above
$\rho\gtrsim0.1$ \gcc\ because of the growing
importance of nonideal contributions and the
onset of pressure ionization. This latter phenomenon is difficult to treat in
the framework of the ``chemical picture'' of plasmas, which
assumes that the different ion species retain their identity
(see, e.g., Refs.\ \cite{SCVH,P96,Rogers00}, for discussions). On
the other hand, EOS calculations within the more rigorous
``physical picture'', quite successful at  relatively low $\rho$
(e.g., \cite{OPAL-EOS}), become prohibitively complicated  at
high densities. First principle approaches based on path
integral Monte Carlo (PIMC) \cite{MC00,Bezkr-ea04} or molecular
dynamics (MD) calculations \cite{SBY01,MazevetEtal2003} are computationally
highly expensive, 
especially at high temperatures when
excited ionic cores should be considered.
To the best of our knowledge, no PIMC or MD
data for carbon or oxygen in the temperature-density range of 
interest in this paper has been published so far.

Besides these first principle based computations and
a few works that consider the influence of
anisotropic distributions of neighboring ions
(e.g. Refs.~\cite{Nardi1990,Wilson2011}), hot dense plasma studies
of medium- to high-$Z$ elements
are generally built around the modeling of an ion in its
plasma environment placing it at the center of a spherically
symmetric system.
The complexity of the physics that is still required leads 
to a wealth of models. 
In ion sphere (IS) models \cite{Lib1979} neighboring ions act as a mere neutralizing background beyond some radius,
while some correlation with the central ion can be introduced
through the computation of pair distribution functions \cite{BF2004},
for instance in the hypernetted chain approximation (HNC) \cite{WVGG2009}.
The free electron background may be obtained from
model prescriptions like Thomas-Fermi \cite{FMT1949,KS95} or may involve quantum 
computations, usually from the Kohn-Sham equations 
in the density functional theory (DFT) context \cite{PBC2009,PCHU2010}.
Several levels of refinement are possible to model the atomic structure
of the central ion. Most of them merge the 
various excitation and/or ionization states 
into a fictitious average atom (AA) \cite{Rozs1972}. This saves a lot of 
computation time because the overall self-consistent scheme for the ion and
its environment has to be solved only once for a given
thermodynamics condition. 
The price to pay, especially when interested in opacities, 
is some additional statistical procedure 
to infer individual ionization stages populations from the 
average-atom solution (e.g. \cite{PDB2006}).
This undoubtedly works well for high-$Z$ elements due to their 
innumerable  quantum states which translate into 
unresolved transition arrays (UTA), but 
for lighter elements such as carbon or oxygen the relevance of
the AA scheme becomes questionable.

In this paper we employ a generalized version of the EOS model
\cite{PMC05} which relies on the free energy minimization in the
framework of the chemical picture and is applied to high densities
across the pressure ionization region.   
We combine separate  models for different ionization
stages in the plasma environment, taking into account the
detailed structure of bound states (configurations, $LS$ terms),
and use Boltzmann statistics to sum up the internal partition
functions of these ions. 
The atomic structure of each ion  embedded
in the dense plasma is calculated using a scheme
\citep{Massacrier} which self-consistently 
takes
into
account the  modification of bound states due to the environment.
The free electron density around each ion is treated
quantum mechanically, thus resolving the resonances.
Though neighboring ions act at this level as a neutralizing background,
the long-range interactions in the system of charged particles (ions
and electrons) is included in the thermodynamics
using the theory previously developed for fully
ionized plasmas  (see Ref.~\citep{PC10} and references therein).
This model allows us to
obtain not only the thermodynamic functions, but also directly number
fractions for every ionization stage, unlike AA models.

As different ions are treated on an equal footing, whether from 
the same element or not, our approach allows to treat mixtures in a 
similar way as pure plasmas.
We apply the model to carbon and oxygen plasmas at temperatures
between $3\times10^5$~K and $3\times10^6$~K and mass densities in the
range $10^{-3}$ \gcc\ $\lesssim\rho\lesssim10^4$ \gcc. We also
consider carbon-oxygen mixtures at the same plasma parameter values and
propose a simple but accurate method of calculation of
thermodynamic functions of the mixtures based on the solution of
the ionization equilibrium problem for the pure substances. At
$T\lesssim3\times10^5$~K the model remains valid in principle, but the
calculation becomes 
numerically
much more difficult
in the current implementation, and it has not been
realized in this work.
For temperatures higher than $\simeq 3\times10^6$ K our model recovers a simpler one
where the free electron density is assumed to be uniform.

In Sec.~\ref{sect:fren} we briefly describe the total free energy
model. In Sec.~\ref{sect:Fint} we present the model for internal
free energy of ions, which takes into account bound state
configurations 
in $LS$ coupling
and their interactions with the continuum of free
electrons. The technique for the calculation of thermodynamic
functions  at equilibrium and its update with respect to
Ref.~\cite{PMC05} is briefly described in Sec.~\ref{sect:TDE}. In
Sec.~\ref{sect:res} we discuss the results of the EOS
calculations for carbon and oxygen plasmas and for their
mixtures. Section~\ref{sect:concl} is devoted to the conclusion.

\section{Free energy model}
\label{sect:fren}

Consider a plasma  consisting of $N_e$ free electrons and
$\Nion=\sum_{j\nu} N_{j\nu}$ heavy ions  in a volume $V$,
where  $N_{j\nu}$ is the number of ions of
the $j$th chemical element having $\nu$ bound electrons,
and $\nu$ can range from 0 to $Z_j$, where $Z_j$ is the
$j$th element charge number. The free
energy model is basically the same as in Ref.~\cite{PMC05}. The
total Helmholtz free energy is
$
    {\Ftot} = F_{e}
    +F_{i}
     + F_\mathrm{ex},
$
where $F_{i,e}$  denotes the ideal free energy of ions and free
electrons, respectively,  and $F_\mathrm{ex}$ is the excess
(nonideal) part, which arises from interactions. The term $F_{i}$
is the 
kinetic
free energy of an ideal Boltzmann gas mixture, which can
be written as
\beq
   F_{i} = \kB T \sum_j \sum_{\nu=0}^{Z_j}
      N_{j\nu} \Big[\ln(\lambda_j^3 N_{j\nu}/V) - 1
       \Big],
\eeq
where $\lambda_j = (2\pi\hbar^2/m_j\kB T)^{1/2}$  is the thermal
de Broglie
wavelength of the ions of the $j$th chemical element in the
plasma, $m_j$ is the mass of these ions, and $\kB$ is Boltzmann
constant. For the electrons at arbitrary degeneracy, $F_{e}$ can
be expressed through Fermi-Dirac integrals (we calculate $F_{e}$
and its derivatives using the code described in \cite{PC10}). The
nonideal term can be written as
\beq
   F_\mathrm{ex} = \FC + F_\mathrm{int},
\label{Fex}
\eeq
where the first term, $\FC$, includes contributions due to the
long-range 
part of the Coulomb interactions between different (classical)
ions, between
free electrons (including exchange and correlations), 
and  between ions and free electrons. The second term, $F_\mathrm{int}$,
linked to internal partition functions,
involves sums over
localized bound states around the nuclei and includes interactions
between bound
and free electrons. 
No strict definition of either free and bound electrons or ions
exists in a dense plasma; therefore, in general, the terms in
\req{Fex} are interdependent. In our approach, we handle this difficulty
as follows: we first calculate the properties of the ions, treated individually, 
embedded in the plasma, by developing self-consistent models for these ions, 
we then couple these models with a model which describes the long
range interactions, and finally we minimize the resulting total free energy
$\Ftot$.

We calculate the Coulomb term and its thermodynamic derivatives
using previously published fitting formulae (see Ref.~\cite{PC10}
for references). Compared to our previous paper \cite{PMC05},
there are two main improvements in the calculation of this term.
First, we employ a correction to the linear mixing rule for the
ions of different types, derived in \cite{PCCDR09}. Second, we
have implemented fully analytical calculations of all derivatives of
$\FC$ needed to obtain thermodynamic functions (previously some
derivations were made numerically). The latter improvement
increases the accuracy of the calculations of these derivatives and
allows us to proceed straightforwardly, with no need to
include the additional refinement described Sec.~IIIC of
Ref.~\cite{PMC05}, based on an extraction of the
long-distance contribution $\FC$ from calculated values of
$\Ftot$.

\section{Bound-state contribution to the free energy}
\label{sect:Fint}

In order to evaluate $F_\mathrm{int}$,
 we calculate the ionic structure in the plasma using the
scheme described in \cite{Massacrier}. It is based on the
ion-sphere approximation, which replaces the actual plasma environment 
for every ion by the statistically averaged plasma effects on
the electron wave functions within a spherical volume centered at
the ionic nucleus. 
At present we do not include neutral atoms ($\nu=Z_j$ for 
the $j$th element), 
which is justified at the temperatures and densities
where the ionization degree of the plasma is high.
For each ion
containing $\nu$ bound electrons, 
a radius of the ion sphere $R_{j\nu}$
is determined self-consistently from the requirement
that the sphere is overall electrically neutral. 
The Hamiltonian describing the ion $(j,\nu)$ immersed in the
plasma is written as
\beq
   H_{j\nu} = \sum_{i=1}^\nu h_{j\nu}(\bm{r}_i)+W_{j\nu},
\label{H}
\eeq
where 
\bea\hspace*{-2em}&&
   h_{j\nu}(\bm{r}) =
   -\frac{\hbar^2}{2m_{e}}\nabla^2+V_\mathrm{at}^{j\nu}(r)+V_f^{j\nu}(r),
\label{h-nu}
\\\hspace*{-2em}&&
   W_{j\nu} =
   \sum_{i=1}^\nu\left(-\,\frac{Z_j
   e^2}{r_i}- V_\mathrm{at}^{j\nu}(r_i)\right)
     + \sum_{i<k}^\nu\frac{e^2}{|\bm{r}_i-\bm{r}_k|},
\label{W-nu}
\eea
$V_f^{j\nu}$ is the potential due to the plasma on the ion
$(j,\nu)$,
that must be determined self-consistently, and
$V_\mathrm{at}^{j\nu}$ is a scaled Thomas-Fermi potential of the
nucleus and 
$\nu$ bound electrons \citep{EN69}
independent of the density and temperature. 
Note that $V_\mathrm{at}^{j\nu}$
disappears in \req{H}.
The effective
Hamiltonian $h_{j\nu}$ generates a one-electron wave function
basis with a finite number of bound states. The coordinate parts
$\psi_{nlm}^{j\nu}$
of the 
bound-state wave
functions are obtained from the
Schr\"odinger equation 
\beq
 h_{j\nu}\psi_{nlm}^{j\nu}
  =\epsilon_{j\nu nl}\psi_{nlm}^{j\nu},
\label{one-el}
\eeq
where $n,l,m$ are, respectively, the principal, orbital, and magnetic 
quantum numbers for a given orbital.
Then $H_{j\nu}$ is diagonalized in 
the
subspace of Slater determinants
generated 
by the set of all $\psi_{nlm}^{j\nu}$. 
The interaction term
$W_{j\nu}$ is
responsible for the $LS$ splitting of configurations.
It may also lead to some configuration interactions.
As  $W_{j\nu}$ does not depend on the plasma properties, its matrix elements
are influenced only through the modifications of the wave
functions in \req{one-el}
due to plasma effects through $V_f^{j\nu}$.
Our experience shows that
the $\nu$-electron energies of the bound states of the
ions are  well approximated as
 $
   E_{j\nu\alpha} = E_{j\nu\alpha}^0 + \sum_{(nl)\in\alpha}
   (\epsilon_{j\nu nl}-\epsilon_{j\nu nl}^0),
 $
where $E_{j\nu\alpha}^0$ and $\epsilon_{j\nu nl}^0$ are calculated
for the isolated ion, and  $\alpha=(nl)_1
(nl)_2\ldots(nl)_\nu\,^{2S+1}L$ defines a particular $LS$ term of
a configuration. The boundary condition at 
the ion sphere radius
$R_{j\nu}$ to solve
\req{one-el} does not noticeably affect $E_{j\nu\alpha}$ except
near the densities where the corresponding term $\alpha$ becomes
pressure-ionized.  At these densities we use the dependence of
the one-electron eigenenergy on the external boundary condition
in order to estimate the degree of electron delocalization due to
pressure effects and to determine the corresponding occupation
probabilities, in the same manner as in Ref.~\cite{PMC05}.

\begin{figure}[t]
\includegraphics[width=\columnwidth]{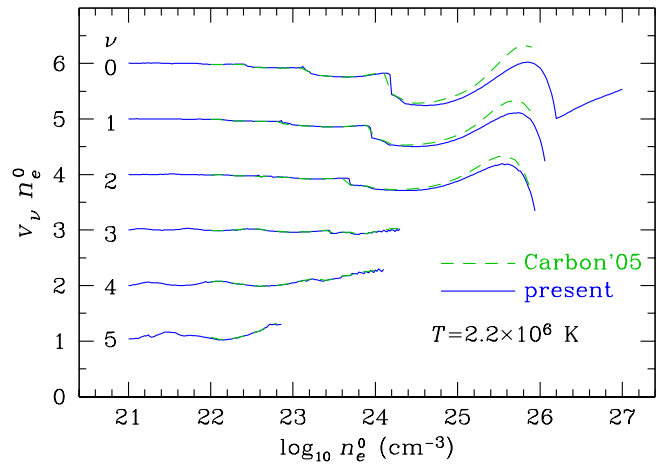}
\caption{(Color online) Neutrality volumes $v_\nu$ normalized with
the fiducial number density of free electrons $n_e^0$ (see text),
as functions of $n_e^0$ for $T=2.2\times10^6$~K
for carbon ions with $\nu$
bound electrons ($\nu=0,1,\ldots,5$).
 The results from
 Ref.~\cite{PMC05} (dashed lines) are compared with
 the present improved results (solid lines).
\label{fig:volC}
}
\end{figure}

The free electron density $n_f^{j\nu}(r)$ and the potential
$V_f^{j\nu}(r)$ are determined self-consistently
in the potential generated by the nucleus charge and a Boltzmann
average of the $\nu$-electron wave-functions, solving for
Kohn-Sham states in the local
density approximation of the DFT, and
assuming Fermi distribution of the free electrons at a given
chemical potential $\mu_e$.
The contributions from
resonances are taken into account, as explained 
in Refs.~\cite{Massacrier,PMC05}. Together with
$n_f^{j\nu}$ and $V_f^{j\nu}$, the ion sphere radii $R_{j\nu}$ and the
corresponding \emph{neutrality volumes} $v_{j\nu}=4\pi
R_{j\nu}^3/3$ are obtained from the neutrality condition for each
ion. 

\begin{figure}[t]
\includegraphics[width=\columnwidth]{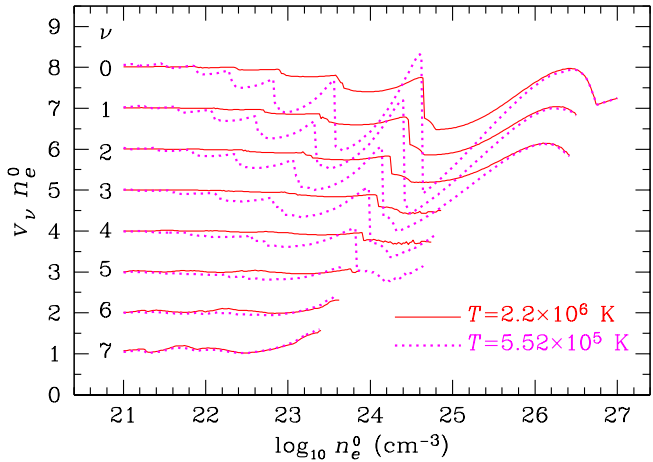}
\caption{(Color online) Neutrality volumes $v_\nu$ normalized with
 $n_e^0$ (analogous
to Fig.~\ref{fig:volC}),
as functions of $n_e^0$ for $T=2.2\times10^6$~K (solid lines)
and $T=5.52\times10^5$~K (dotted lines) for oxygen ions with $\nu$
bound electrons ($\nu=0,1,\ldots,7$).
\label{fig:volO}
}
\end{figure}

For a uniform free
electron background, we would have
$v_{j\nu} = v_{j\nu}^0 = (Z_j-\nu)/n_e^0$, where 
$n_e^0(\mu_e,T)$ is the ``fiducial'' electron density, which
would correspond to the true electron density of the ideal Fermi
gas of electrons at the given $\mu_e$ and $T$ values. When taking into account
the interactions of the free electrons with the ions,
$v_{j\nu}$ deviates from $v_{j\nu}^0$, as illustrated in
Figs.~\ref{fig:volC} and \ref{fig:volO} for carbon and oxygen
plasmas, respectively. The drops of the curves at certain
values of $n_e^0(\mu_e,T)$, which are especially sharp at
lower temperatures (see Fig.~\ref{fig:volO}), are the consequence
of pressure ionization of separate bound levels: when a $nl$ level of
ion $\nu+1$  crosses the continuum limit and appears as a
resonance in the neighboring ionization state $\nu$, the corresponding
ion sphere shrinks to compensate this increase in the free
electron density of states (note that $Z-\nu$ free electrons are
still present in the volume $v_{j\nu}$).
For a given level $nl$, pressure ionization occurs at electronic densities 
which roughly depend on the ionic charge as $(Z_j-\nu)^4$.
If the number of 
bound states
that remain in the ion sphere is not sufficient to accept $\nu$
electrons, that ion disappears: $\nu=5,6,7$ disappear with the $2p$ sub-shell,
and $\nu=3,4$ with $2s$, see Figs.~\ref{fig:volC} and \ref{fig:volO}.
We emphasize that, in contrast to average-atom models, separate ions
have their own fate.
Compared to Ref.~\cite{PMC05}, we have
extended the calculation of the free electron states to higher
electron energies and to higher $\mu_e$. The consequences of
these improvements for the neutrality volumes are seen in
Fig.~\ref{fig:volC}, where the dashed lines show our previous results
and the solid lines correspond to our new calculations. The
extension of $\mu_e$ to higher values allows us to capture the
last dips of the neutrality volumes of naked nuclei ($\nu=0$) as
functions of $n_e^0$, which appear at $n_e^0>10^{26}$ cm$^{-3}$
for both carbon and oxygen. These dips are due to the coupling of the
hydrogenlike ($\nu=1$) bound states to the continuum, and are
related to the pressure ionization of the K shell (where the
curves $\nu=1$ in Figs.~\ref{fig:volC} and \ref{fig:volO}
terminate).

Pressure ionization of each atomic shell is accompanied by a
spreading of the electronic level into a band, eventually passing into the
continuum. We account for this effect by using occupation
probabilities 
$w_{j\nu\alpha} = \prod_{(nl)\in\alpha} w_{j\nu nl}$, 
which are equal to the fraction
of the band corresponding to the electron configuration $\alpha$
that remains below the continuum level 
at given $(T,\mu_e)$ values
(see Ref.~\cite{PMC05} for details). 

In order to get a smooth pressure ionization and a smooth
extension from calculated values of the neutrality volumes to
higher densities,  we replace $v_{j\nu}$ by
$v_{j\nu}^\ast=v_{j\nu} w_{j\nu}+v_{j\nu}^0 (1-w_{j\nu})$, where
$w_{j\nu}$ is the occupation probability $w_{j\nu\alpha}$
calculated for the ground-state configuration $\alpha$ of the ion
with $\nu$ electrons.

The separation of $H_\nu$ into parts (\ref{h-nu}) and
(\ref{W-nu}) allows us to capture the plasma effects in
one-electron energies and wave functions through
Eqs.~(\ref{h-nu}) and (\ref{one-el}),
while the $\nu$-electron structure -- configuration energy mean shifts
and $LS$ splitting -- results from the contribution $W_\nu$.
In Ref.~\cite{PMC05} we calculated this
contribution to binding energies using 
a modified version of
the ``Superstructure'' code
\cite{EN69}, as in Ref.~\cite{Massacrier}. But our experience
has revealed that the use of the atomic code is unnecessary
for our purposes. This contribution is calculated once 
for the isolated ion
and then
combined with
 the one-electron energies 
$\epsilon_{j\nu nl}$ (see above) at any density.
Indeed the dependence of
this contribution
on the plasma conditions  
 is weak (of the second order in
perturbation) compared 
to the dependence of the energies
$\epsilon_{j\nu nl}$. In the present work we
have updated the corrections due to  $W_\nu$ by using the
detailed atomic energy database of the Opacity Project (OP)
(Ref.~\cite{Seaton05} and references therein).

Having calculated energies $E_{j\nu\alpha}$ and occupation
probabilities $w_{j\nu\alpha}$
at given values $(T,\mu_e)$, we
obtain the contribution of the internal degrees of freedom 
of all the ions to
the free energy as
\beq
   F_\mathrm{int} = - \Nion \kB T \sum_j \sum_{\nu=0}^{Z_j}
    x_{j\nu} \ln\mathcal{Z}_{j\nu},
\eeq
where 
$
   \mathcal{Z}_{j\nu} = \sum_{\alpha} 
   w_{j\nu\alpha}\,(2S+1)(2L+1)\,\exp(-E_{j\nu\alpha}/\kB T)
$
is the internal partition function of the
ion $(j,\nu)$ in the plasma, and
$x_{j\nu} \equiv N_{j\nu}/\Nion$ its number fraction ($\sum_{j,\nu}
x_{j\nu} = 1$). 
Since we do not consider neutral atoms, the
actual upper limit of  $\nu$ summation is $Z_j-1$.

\section{Thermodynamic equilibrium}
\label{sect:TDE}

We use a generalization of the free energy minimization method of
Ref.~\cite{PMC05} to the case of multicomponent mixtures. The
minimum of $\Ftot$ is sought at constant 
$V$, $T$, and $N_j$. The free
parameters are the chemical potential $\mu_e$ and the ion
fractions $x_{j\nu}$, subject to constraints
\beq
   x_{j\nu} \geqslant 0, 
   \quad
   \sum_{\nu=0}^{Z_j-1} x_{j\nu} = Y_j,
   \quad
   \Nion \sum_j \sum_{\nu=1}^{Z_j-1} x_{j\nu} v_{j\nu} = V,
\label{eq:min}
\eeq
where $Y_j$ are fixed chemical element abundances ($Y_j \geqslant
0$, $\sum_j Y_j = 1$). The last condition in \req{eq:min}
reflects that $V$ is constant. We ensure this
condition by using the Lagrange multiplier method. The other
conditions in \req{eq:min} are imposed explicitly in our
minimization procedure by allowing the set of $\{x_{j\nu}\}$ to
contain only the values for which these constraints are satisfied.
Note that the charge neutrality condition is fulfilled
automatically, because the volume $v_{j\nu}$ around every nucleus
is neutral by construction. Some further details of the
minimization algorithm can be found in Ref.~\cite{PMC05}. For a
mixture of chemical elements, we have in total $1+\sum_j (Z_j
-1)$ independent parameters of minimization.

The minimization procedure provides the values of $\Ftot$ with a
typical accuracy of four digits. Further improvement of the
accuracy becomes problematic. The main sources of the numerical noise
are the finite precision of calculation of the internal partition
functions and neutrality volumes and the finite precision of the
minimization procedure. The achieved accuracy is not sufficient
for an accurate evaluation of thermodynamic functions, especially
those that involve second and mixed derivatives of $\Ftot$
(specific heat $C_V$, logarithmic pressure derivatives with
respect to density, $\chi_\rho$, and temperature, $\chi_T$,
adiabatic temperature gradient, and so on). To overcome this
difficulty, we use an additional filtering of the numerical noise
after the minimization.  The filtering procedure is modified
relative to that in Ref.~\cite{PMC05}. Specifically, for the
results shown in the next section, we have performed calculations for
a grid of points $(\rho,T)$ separated by $\Delta\log_{10}\rho=0.05$
and $\Delta\log_{10} T=0.1$, and at each point we have determined an improved
value of $\Ftot$ using a bicubic 10-parameter polynomial fit to
$\Ftot(\rho,T)$ on this grid. The fitting was done by $\chi^2$
minimization with weights decreasing with increasing distance
between a given point $(\rho,T)$ and the grid point $(\rho',T')$
in the $\log_{10}\rho$--$\log_{10} T$ plane as
$[1+(\log_{10}(\rho'/\rho)/\Delta\log_{10}\rho)^2 +(\log_{10}(T'/T)/\Delta\log_{10}
T)^2]^{-1}$.  This improved method of primary filtering of
numerical noise allows us to avoid the additional secondary
smoothing of thermodynamic functions that was employed in
Ref.~\cite{PMC05}.

\section{Results}
\label{sect:res}

\subsection{Ionization equilibrium}
\label{sect:ioneq}

\begin{figure}[t]
\includegraphics[width=\columnwidth]{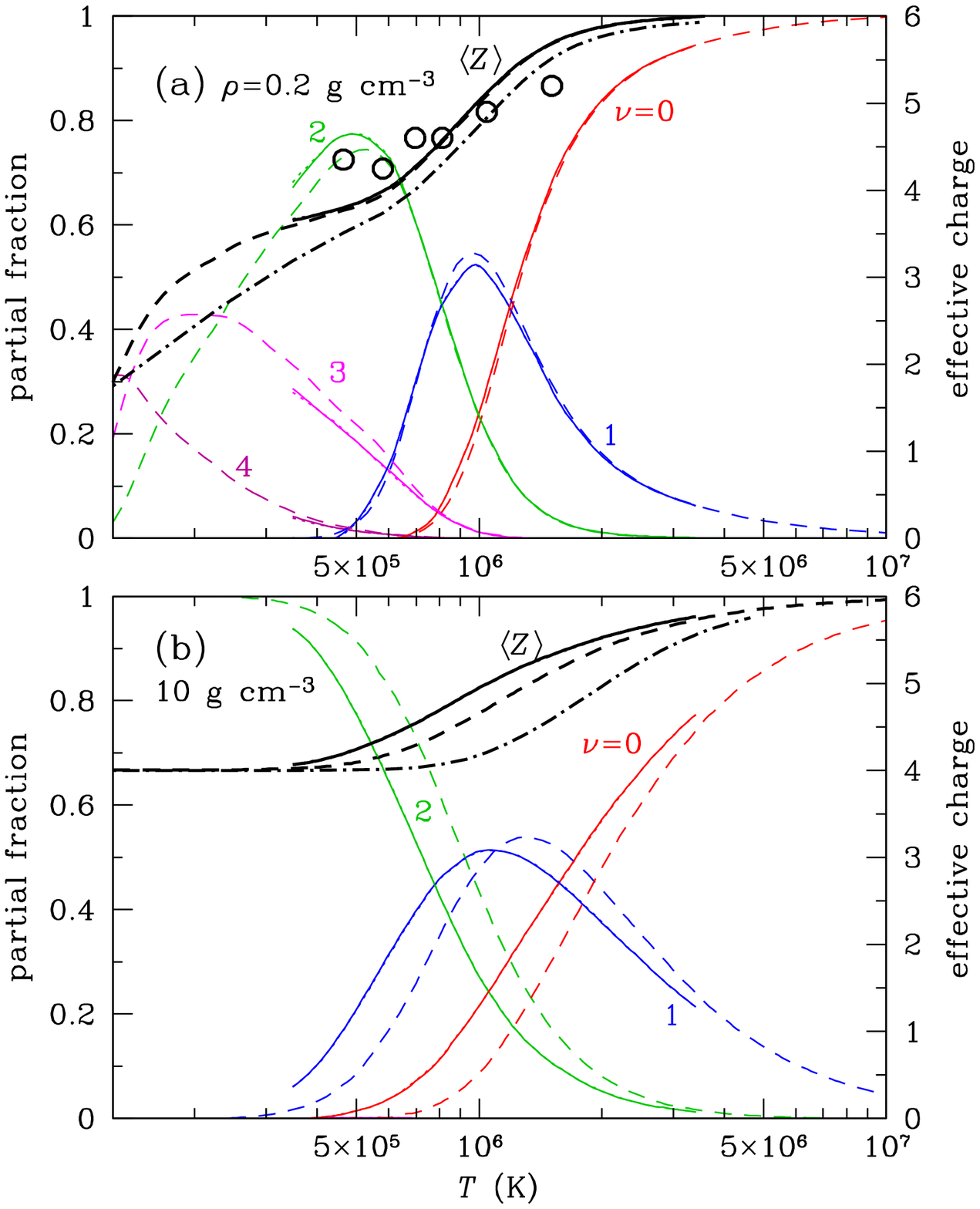}
\caption{(Color online) 
Number fractions of different carbon ions (left vertical axis,
marked by the numbers of bound electrons $\nu$ near the curves) 
and mean effective charge $\langle Z\rangle$ (right vertical axis)
in a pure plasma versus temperature
at  (a) $\rho=0.2$ \gcc\ and (b) $\rho=10$ \gcc. 
Solid lines: accurate results; dashed lines: approximation
where partition functions
$\mathcal{Z}_{j\nu}$ and neutrality volumes $v_{j\nu}$ are calculated assuming
a uniform free electron density.
Dash-dotted line: results for  $\langle Z\rangle$ from FLYCHK \cite{Chung2005}.
In panel (a) the circles are the experimental results from \citet{Gregori2008}
for  $\rho=0.2$ \gcc.}
\label{fig:Cfrac}
\end{figure}

\begin{figure}[t]
\includegraphics[width=\columnwidth]{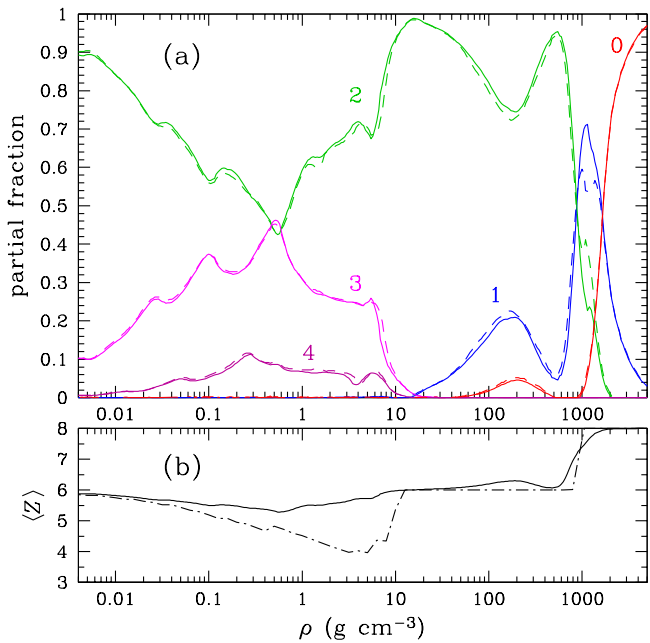}
\caption{(Color online) 
(a) Number fractions of ions for a pure oxygen plasma  (solid lines)
at $T=6\times10^5$~K as a function of the density $\rho$.
Numbers of bound electrons $\nu$ for each ion are marked near the curves.
Dashed lines are \emph{twice} the ionization fractions $x_{j\nu}$ for
oxygen in a plasma mixture composed of equal numbers of C and O nuclei
at the same temperature.
(b) Mean effective charge $\langle Z\rangle$ for the pure oxygen plasma (solid line).
The dot-dashed line is the result from FLYCHK under the same conditions.
\label{fig:Ofrac5t06}
}
\end{figure}

\begin{figure}[t]
\includegraphics[width=\columnwidth]{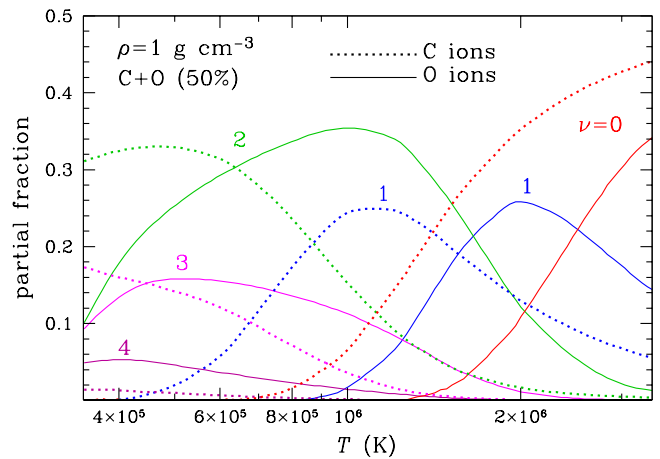}
\caption{(Color online) 
Number fractions of different 
carbon (dotted lines) and oxygen (solid lines)
ions in the mixture of equal number of ions of each
chemical element at $\rho=1$ \gcc\ as functions of $T$.
\label{fig:COfrac}
}
\end{figure}

\begin{figure}[t]
\includegraphics[width=\columnwidth]{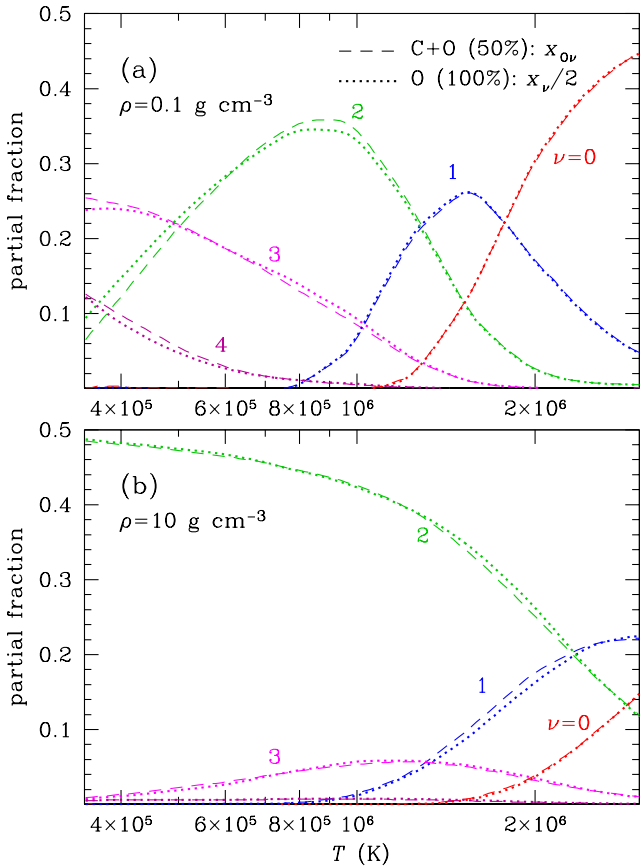}
\caption{(Color online) 
Number fractions of different oxygen ions as functions of $T$
in the mixture of equal number of ions of each
chemical element C and O at (a)
$\rho=0.1$ \gcc\  and (b) $\rho=10$ \gcc.
Dashed lines show the result of direct free energy minimization
for the mixture, and dotted lines show the result of an
approximation based on the solution of the ionization equilibrium
problem for pure C and O plasmas (see text).
\label{fig:Ofracmix}
}
\end{figure}

\begin{figure}[t]
\includegraphics[width=\columnwidth]{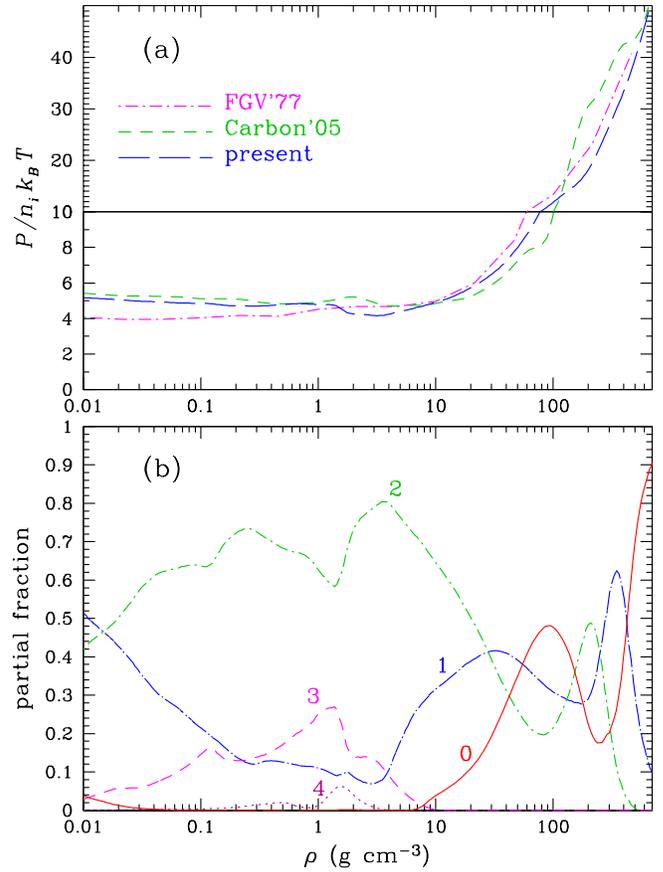}
\caption{(Color online) 
(a) Normalized pressure $PV/\Nion\kB T$ for pure
carbon along the isotherm $T  = 5\times10^5$~K.
The present results (long-dashed line) are compared with the
results from Ref.~\cite{FGV77} (dot-dashed line)  and
Ref.~\cite{PMC05} (short-dashed line). Note the different scale 
in the figure above and below the middle horizontal line.
(b) Number fractions of different carbon ions (present results)
for the same temperature.
\label{fig:P}
}
\end{figure}

\begin{figure}[t]
\includegraphics[width=\columnwidth]{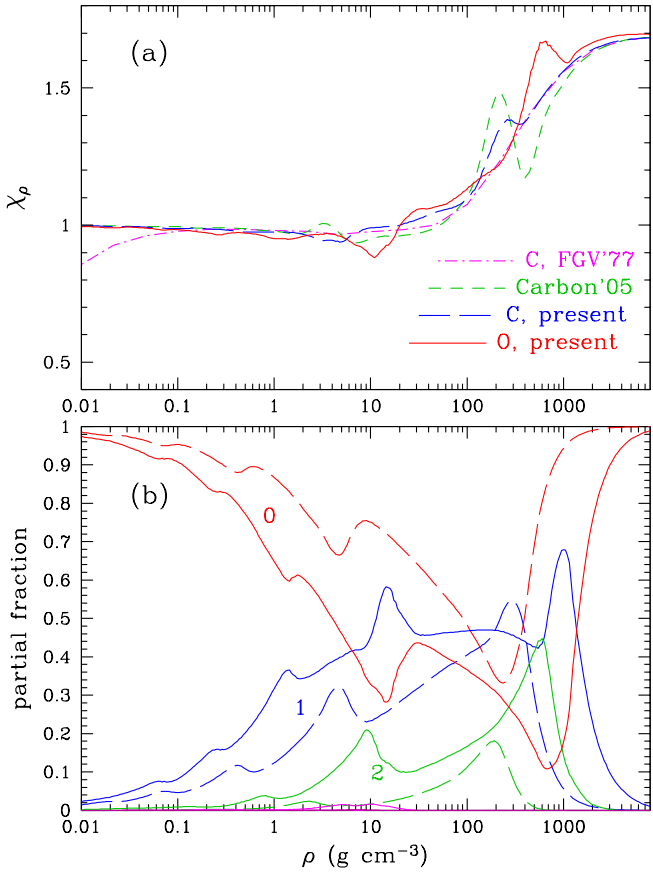}
\caption{(Color online) 
(a) Logarithmic pressure derivative
$\chi_\rho=\partial\ln P/\partial\ln\rho$ along the isotherm 
$T  = 3.16\times10^6$~K.
The present data for carbon (long-dashed line) 
are compared with the results
from Ref.~\cite{FGV77} (dot-dashed line) 
and Ref.~\cite{PMC05} (short-dashed line).
The solid line shows the isotherm of $\chi_\rho$ for oxygen.
(b) Present results for the number fractions of carbon ions (dashed lines)
and oxygen ions (solid lines) at the same temperature.
\label{fig:chi_r}
}
\end{figure}

The minimization of the total free energy immediately gives a
solution to ionization equilibrium. Examples of
partial fractions of ions as functions of $T$ at constant $\rho$,
or of $\rho$ at constant $T$, 
are shown in Figs.~\ref{fig:Cfrac}--\ref{fig:Ofracmix},
as well as in panels (b) of Figs.~\ref{fig:P} and \ref{fig:chi_r} in Sect.~\ref{subsec:thermofct}.
Figure~\ref{fig:Cfrac} shows current results (solid lines) in a pure carbon plasma 
for the partial fractions of ions with different
numbers of electrons (marked by the numbers near the respective
curves)  and for the mean effective charge (marked by the symbol
$\langle Z\rangle$ and scaled to the right vertical axis) at (a) $\rho=0.2$ {\gcc}
and (b) $\rho=10$ \gcc\ . 
This effective charge is calculated by taking into account the
partial delocalization of the outer-shell electrons in the
pressure ionization region as $\langle Z\rangle = \sum_{j\nu}
(Z_j-\nu+1-w_{j\nu})\,x_{j\nu}$. 
The ions recombine with decreasing temperature at both densities,
but at the high density $\rho=10$ \gcc\ there is no room for the existence in carbon of
bound states other than $1s$
(as can be inferred from Fig.~\ref{fig:volC}, though for a fixed temperature: 
$\rho=10$ \gcc\ corresponds to $n_e\simeq~2$\,--\,$3\times10^{24}$~cm$^{-3}$.)
Hence recombination can not proceed past the He-like ($\nu=2$) stage.

The dashed lines in the figure
correspond to a simplified approach, where the  neutrality
volumes $v_{j\nu}$ and internal partition functions
$\mathcal{Z}_{j\nu}$ are calculated assuming the uniform electron
density $n_e=n_e^0$. In this simple approximation we can obtain
results in a wider range of $T$ than in the accurate calculation. 
This approximation works
well at relatively low densities, as proved by the good agreement
of the dashed and solid curves in  Fig.~\ref{fig:Cfrac}(a). With
increasing density, the coupling of the bound and continuum states
becomes significant. 
More accurate calculation becomes necessary, 
as can be seen in Fig.~\ref{fig:Cfrac}(b),
especially at low temperatures where the difference in the ion
sphere between a uniform free electron density and the density constructed from
quantum continuum states will be more pronounced. 

We also report on Fig.~\ref{fig:Cfrac} the mean ion charge as obtained from FLYCHK
(dash-dotted lines),
a suite of codes based on simplified atomic models aimed at providing fast collisional-radiative
models and spectra \cite{Chung2005}. FLYCHK gives generally a lower mean effective charge, especially
at  temperature-density conditions where several ionization stages are present. 
This is probably due to its treatment of density effects through a simplified approach. 
The experimental data points of \citet{Gregori2008} for $\langle Z\rangle$ 
are plotted in Fig.~\ref{fig:Cfrac}(a) as circles. They agree with our results at this density,
especially if one considers the experimental uncertainties of $\pm 20$ eV for the temperature
and $\pm 0.25$ for  $\langle Z\rangle$
\footnote{the FLYCHK mean ion charge $\langle Z\rangle$ reported here for
$\rho=0.2$ {\gcc} and obtained from the NIST website appears to be different from the data
as plotted on Fig.~4 in \citet{Gregori2008}. We think there is an inversion in the naming of the
FLYCHK and PURGATORIO results on that figure in this paper.}.
Experiments at higher densities would certainly be very valuable.


The effects of pressure ionization are most appreciable 
in Fig.~\ref{fig:Ofrac5t06}, where ion population fractions 
(a) and the mean effective charge (b)
for
a pure oxygen plasma (solid lines) are plotted versus density at $T=6\times10^5$~K. 
The results of FLYCHK for  $\langle Z\rangle$ are also reported on Fig.~\ref{fig:Ofrac5t06}(b)
(dot-dashed line).
Recombination with increasing density is modulated at the lowest 
densities ($\rho\lesssim 0.5$ \gcc) by the successive
pressure ionizations of the Li-like ($\nu=3$) ion shells
(see Fig.~\ref{fig:volO}: $\rho= 0.5$ \gcc\ corresponds 
to $n_e\simeq 10^{23}$~cm$^{-3}$ for $\langle Z \rangle=5.5$). 
The temperature is sufficient to populate excited states of that
ion but not of He-like ions, so that only 
the Li-like ion partition function is affected
(the $1s^2~2s$--$1s^2~3s$ energy difference is $9\times 10^5$~K, while
the $1s^2$--$1s~2s$ one is $6.5\times 10^6$~K).
Between $\rho\simeq 0.5$ \gcc\ and $20$ \gcc, Li-like ions gradually disappear
with the $n=2$ shell to the benefit of He-like ions.
On the contrary (see Fig.~\ref{fig:Ofrac5t06}(b)), FLYCHK still allows recombination 
into the Be-like ion ($\nu=4$)
until $\rho\simeq 8$ \gcc\ where it is sharply pressure ionized, just before Li-like ions.
Oxygen remains at higher densities in the He-like stage up to $\rho\simeq 10^3$ \gcc\ 
where  the last $1s$ shells for He-like and H-like ions are moved into the continuum.
In this density range our results show a more complex behaviour due to several effects:
the free electrons become degenerate,
the ionization occurs on a large density range due to band broadening, 
and the influence of neutrality volumes is strong.

Dashed lines in Fig.~\ref{fig:Ofrac5t06}(a) show the ionization fractions
of oxygen in a mixture of equal numbers of C and O nuclei. The fractions $x_{j\nu}$
have been multiplied by two to be compared with the pure oxygen plasma case.
The behavior of the ionization equilibrium of oxygen
is quite similar in the two cases, the
biggest difference occurring in the pressure ionization zone
around  $\rho\simeq 10^3$~\gcc.
In Fig.~\ref{fig:COfrac} we show the temperature
dependence of the partial ion fractions for the two chemical
elements C (dotted lines) and O (solid lines) in the same mixture
at $\rho=1$ \gcc.
At this density, the curves for the fraction of $\nu$-ions for 
carbon are roughly shifted to lower temperatures when compared
to the oxygen ones by the ratio of the ionization energies
$(Z_C-\nu)^2/(Z_O-\nu)^2$.


Comparing  Fig.~\ref{fig:COfrac}
 and analogous figures for
mixtures and pure substances, we notice that the temperature
and density dependences of partial ion fractions in mixtures of
different elements are similar to those in pure substances,
scaled by the element abundances $Y_j$. 
Two examples are presented
in Fig.~\ref{fig:Ofracmix}. Here, the dashed lines show partial
fractions $x_{j\nu}$ 
of oxygen ions with $\nu$ bound electrons in
a mixture where half the ions are oxygen ions and half are carbon ions,
at (a) $\rho=0.1$ \gcc\ and (b) $\rho=10$ \gcc, 
as function of $T$, while the dotted lines
show fractions of the same ions, \emph{multiplied by} 0.5, in a
pure oxygen plasma. The close agreement between these curves,
which has been tested at several densities (see, e.g., Fig.~\ref{fig:Ofrac5t06}(a))
or compositions, 
prompted us
to suggest the following method for an approximate solution of the
ionization equilibrium problem and a construction of the EOS for plasma mixtures of
different elements 
(at least in the temperature range considered in the present study): 
\begin{itemize}
\itemsep=0pt
\item calculate the fractional
numbers of different ions $x_{j\nu}$ for pure substances at the
same $n_e$ and $T$ (let us denote their values $x_{j\nu}^0$);
\item  multiply them by $Y_j$ and
keep them fixed: $x_{j\nu}=x_{j\nu}^0 Y_j$;
\item  adjust the electron chemical potential $\mu_e$ so as to
fulfill the last condition in \req{eq:min}. 
\end{itemize}
The advantage of this
method for a mixture of $J$ chemical elements is that, instead of
minimizing $\Ftot$ in a space of $1+\sum_{j=1}^J (Z_j -1)$
independent parameters, one needs to perform $J$ minimizations in
a space of $Z_j$ parameters each, and then to adjust $\mu_e$.
As a rule, these $J$ partial minimizations go much
faster than the single minimization in the space of all
parameters, thus saving the computer cost of the procedure. For
example, it is much easier to find the minima of $\Ftot$ for
carbon (6 parameters) and for oxygen (8 parameters) separately,
than to find a minimum for their mixture (13 parameters).
Moreover, having found the number fractions of ions for pure
substances once, one can calculate thermodynamic functions of
mixtures with \emph{any} fractions of chemical elements $Y_j$
without repeating the 
whole minimization procedure.
Note however that carbon and oxygen are chemical species that are not
very different, so that one should proceed with caution before applying
this simplified approach to other mixtures without further studies.

\begin{figure}[t]
\includegraphics[width=\columnwidth]{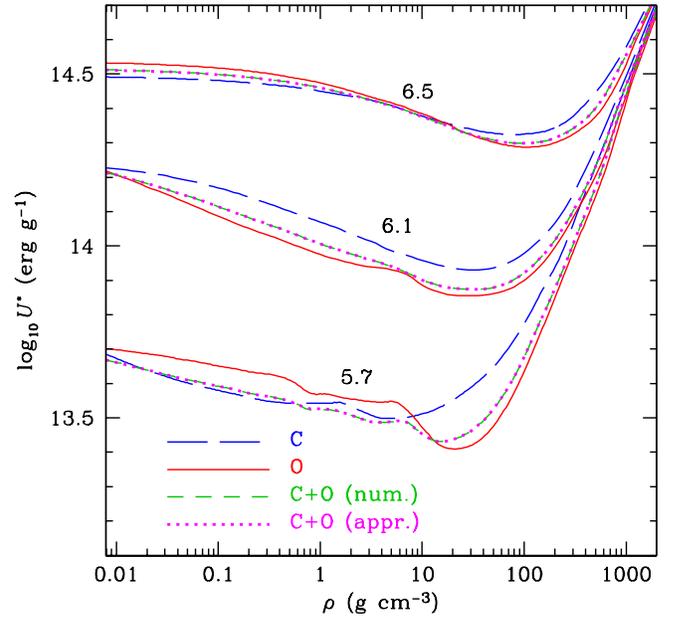}
\caption{(Color online) Specific internal energy logarithms (in cgs
units) for various isotherms:
$T = 5\times10^5$~K, $1.26\times10^6$~K, and $3.16\times10^6$~K
(the curves are marked by $\log_{10}T$ values)
for oxygen (solid lines), carbon (long-dashed lines), and 
the mixture of equal number of ions of each
chemical element (short-dashed and dotted lines).
In the latter case short-dashed lines show the result of 
direct free energy minimization
for the mixture, while dotted lines show the result of an
approximation based on the solution of the ionization equilibrium
problem for pure C and O plasmas (the difference between them
is almost unnoticeable; see text).
\label{fig:U}
}
\end{figure}
\begin{figure}[t]
\includegraphics[width=\columnwidth]{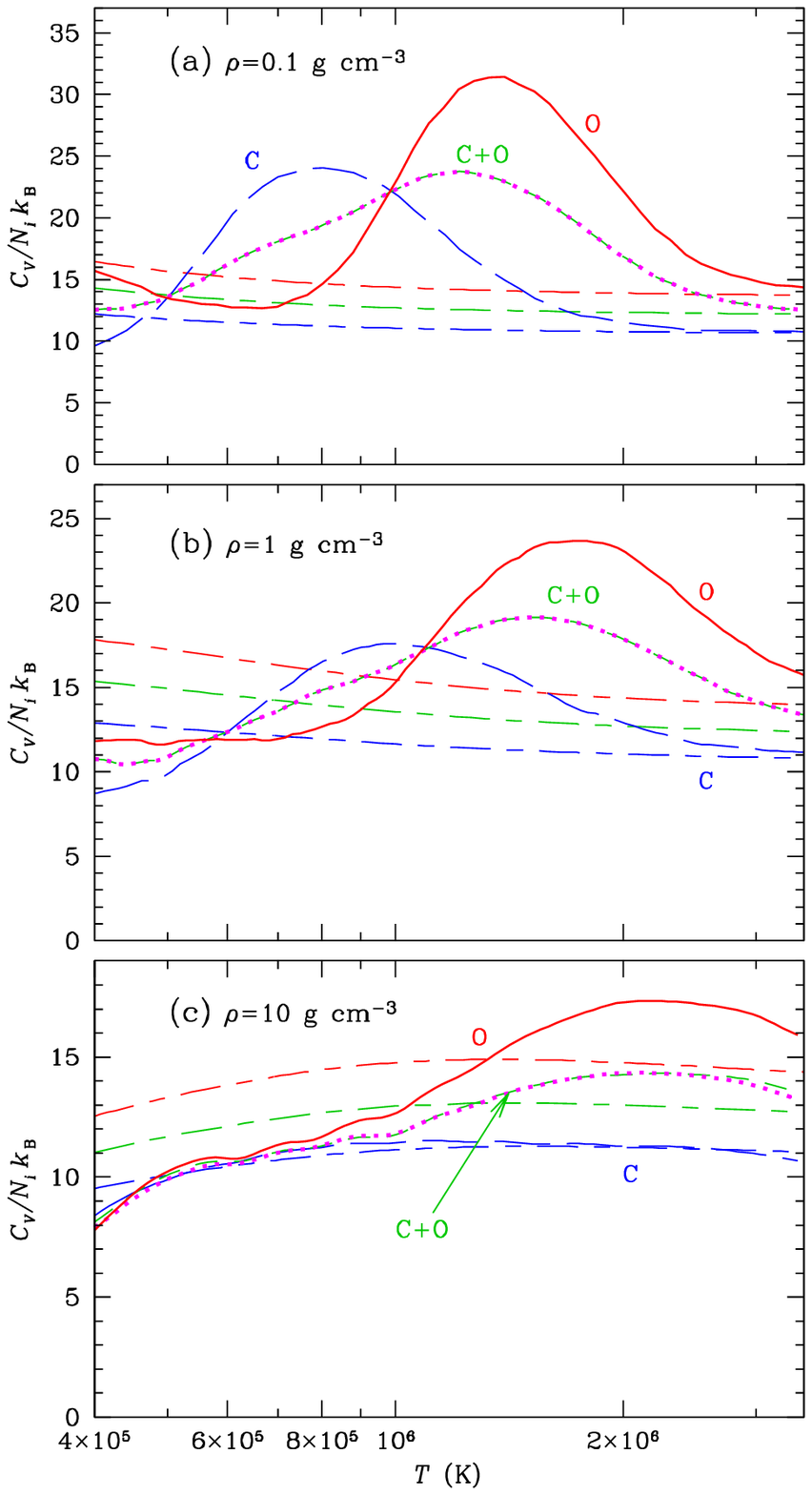}
\caption{(Color online) 
Normalized specific heat per ion for the isochores
(a) $\rho = 0.1$ \gcc\ , (b) $\rho=1$~\gcc\ ,
and (c) $\rho=10$~\gcc\ 
for carbon (long-dashed line), oxygen (solid
line), and mixture of equal number of ions of each
chemical element (short-dashed and dotted lines).
In the latter case, the short-dashed line shows the result of 
direct free energy minimization
for the mixture, while the dotted line shows the result of an
approximation based on the solution of the ionization equilibrium
problem for pure C and O plasmas.
For comparison, alternating short and long dashes
show analogous isochores
for the fully ionized plasma model.
\label{fig:CVs}
}
\end{figure}

\subsection{Thermodynamic functions}
\label{subsec:thermofct}

To illustrate our results for the EOS and compare them with the
results of Ref.~\cite{PMC05}, we show in Fig.~\ref{fig:P}(a) an isotherm for
the pressure $P(\rho)$ for carbon at $T=5\times10^5$~K. Our present result
is shown by the long-dashed line, our previous result \cite{PMC05} by
short dashes, and the dot-dashed line shows the result obtained with the FGV EOS
\cite{FGV77} based on a Thomas-Fermi model at $\rho\gtrsim3$
\gcc. In order to make the differences visible in the figure, we
have normalized the pressure by its value for an
ideal Boltzmann gas of ions, $\Nion\kB T/V$. The vertical scale
is smaller for the upper part of the figure, to take into account
the rapidly increasing pressure contribution of degenerate electrons. The
differences between our and FGV results have become smaller
for $\rho\gtrsim10$ \gcc\ 
 than
in Ref.~\cite{PMC05}, and the additional features have become
less spectacular. This change is mainly caused by the improved
treatment of the electron continuum in the present work (see
Sec.~\ref{sect:fren}).

However, the additional features still persist near the densities where electron shells become
pressure ionized, as expected. 
For instance it can be seen from the ionization fractions as given in 
Fig.~\ref{fig:P}(b) for the same isotherm
that the feature in the pressure just above $\rho=1$~\gcc\ 
is linked to the disappearance of the Li-like carbon ion.
These features are clearly revealed in the
logarithmic derivative of the pressure, $\chi_\rho=(\partial\ln
P/\partial\ln\rho)_T$, shown in Fig.~\ref{fig:chi_r}(a). In this
panel, the dot-dashed, short-dashed, and long-dashed lines
correspond to the same three models for carbon plasma as in
Fig.~\ref{fig:P}(a). In addition, the solid line shows
$\chi_\rho$ for the oxygen plasma at the same 
temperature. 
When analyzed with the help of  the ionization fractions as plotted in Fig.~\ref{fig:chi_r}(b) 
the features for carbon around  $\rho\simeq 5$~\gcc\ and  $3\times10^2$~\gcc\ appear to be linked
to the pressure ionization in H-like ions of the $n=2$ and $n=1$ shells respectively;
those for oxygen around $\rho\simeq 10$~\gcc\ and  $6\times10^2$~\gcc\ are linked 
to the same shells with a supplementary contribution from the He-like ions.
In the latter case, the features due to pressure
ionization are more pronounced because of the higher binding
energies compared to carbon, and they are shifted to higher densities
because of the higher net charge and respectively smaller
neutrality volumes of ions (this shift can be
roughly estimated as the ratio of the 
volumes of hydrogen-like carbon and oxygen ions,
$\sim(Z_O/Z_C)^3=64/27$). 
Note also that the pressure ionization of a given (sub-)shell leaves an imprint
on thermodynamics functions at different densities depending on which ionization stage
is dominant. For instance the features due to the $n=2$ shell in carbon appear at 
$\rho\simeq 1$~\gcc\  or $5$~\gcc\ depending on whether Li-like or H-like ions are
concerned (compare Fig.~\ref{fig:P} and Fig.~\ref{fig:chi_r}).

Figure \ref{fig:U} shows three isotherms for the internal energy
per unit mass for carbon (long-dashed lines), oxygen (solid
lines), and the mixture containing equal numbers of C and O
nuclei (short-dashed and dotted lines). The energy per unit 
mass
$U^*=U/(\Nion\sum_j m_j Y_j)+U_0^*$  
is measured from the ground
state  energy of nonionized atoms, which corresponds to  a shift
with respect to the electron continuum level equal to
$U_0^*=\sum_j Y_j U_{0,j}^*$, where $U_{0,j}^*=8.28\times10^{13}$ erg
g$^{-1}$ for carbon and $U_{0,j}^*=1.23\times10^{14}$ erg g$^{-1}$
for oxygen. 
For the mixture, the short-dashed lines
portray the results of the numerical minimization of the complete
free energy $\Ftot$, while the dotted lines illustrate the results
of the approximate method described in Sec.~\ref{sect:ioneq}.
The near coincidence of the dotted and short-dashed lines,
indistinguishable at the scale of the figure, confirms the
accuracy of the approximate method in that case.

As an example of a second-order thermodynamic function calculated
with our model, Fig.~\ref{fig:CVs} presents
isochores for the heat capacity at (a) $\rho=0.1$ \gcc\ ,
(b) $\rho=1$~\gcc\ , and 
(c) $\rho=10$~\gcc\  for carbon
(long-dashed line), oxygen (solid line), and the carbon-oxygen
mixture with $Y_j=0.5$.
For the mixture, the short-dashed lines
portray the results of the numerical minimization of the complete
free energy $\Ftot$, while the dotted lines illustrate the results
of the approximate method described in Sec.~\ref{sect:ioneq}.
As well as in Fig.~\ref{fig:U}, the short-dashed and dotted
lines nearly coincide.
For comparison, lines drawn with 
alternating short and long dashes show the results
obtained for the fully ionized
plasma model. The presence of bound states yields
big bumps on the curves. 
Comparison with Figs.~\ref{fig:Cfrac} and \ref{fig:Ofracmix}  (for $0.1-0.2$ \gcc\ and $10$ \gcc)
and with Fig.~\ref{fig:COfrac}  (for $1$ \gcc) 
reveals that these
bumps are associated with the thermal ionization of heliumlike ions into
hydrogenlike ones and then into nuclei.
At higher temperatures, the
results for fully ionized plasmas are recovered,
as it should be due to the complete ionization of C and O
at such high temperatures, in agreement with the Saha equation. 

\section{Conclusions}
\label{sect:concl}

We have further improved our model \cite{PMC05} for the calculation
of the EOS for  dense, partially ionized plasmas, based on the
free energy minimization method, suitable to handle pressure
ionization regimes. The free energy model is constructed in the
framework of the chemical picture of plasmas and includes
a detailed, self-consistent treatment of the quantum states of
partially ionized atoms in the plasma environment
as well as a quantum treatment of the continuum electrons.  The
improvements with respect to our former pure carbon calculations include implementing updated fitting formulae
\cite{PCCDR09,PC10} for the long-range Coulomb contribution to
the free energy, the use of the OP database for energies of bound
state configurations \cite{Seaton05}, an increased range of density
and the extension to mixtures of different chemical
elements.

We have extended 
the calculations from pure carbon to arbitrary
carbon-oxygen mixtures. We
have also suggested an efficient, approximate but accurate method
of EOS calculation in the case of mixtures of different chemical
elements, based on the detailed information about the various
ionic state fractions for the pure species.

The present EOS results can be used in studies of
carbon-oxygen plasmas in the domain of so-called warm dense matter 
as well as in
astrophysical calculations of stellar structure and evolution.
The possible astrophysical applications include
oxygen and carbon
plasmas and carbon-oxygen mixtures
in various types of stars, for example
the interiors of carbon-oxygen white dwarfs.
However, in order to apply it also to the outer
parts of the white dwarf envelopes, it is desirable to extend
the results to lower temperatures. We are planning
to perform such extensions in the future work.

\begin{acknowledgments}
We thank an anonymous referee for having brought to our attention
the paper by \citet{Gregori2008}, and
Y. Ralchenko for having provided us access to FLYCHK on the NIST website.
G.~M. acknowledges partial support from the 
Programme National de Physique Stellaire (PNPS) of CNRS/INSU.
The work of A.Y.P.\ was supported in part by
the Russian Foundation for Basic Research (RFBR Grant 11-02-00253-a)
and Rosnauka ``Leading Scientific Schools'' Grant NSh-3769.2010.2. 
The research leading to these results has received partial funding from the European Research Council under the European Community's Seventh Framework Programme (FP7/2007-2013 Grant Agreement no. 247060).
\end{acknowledgments}


\end{document}